\documentclass[aps,prl,letterpaper,twocolumn,superscriptaddress,showpacs,amsmath,amssymb]{revtex4-1}
\usepackage{graphicx}  % needed for figures
\graphicspath{ {figures/} }
\usepackage{bm}     % for math
\usepackage{amssymb}   % for math
\usepackage{amsmath}   % for math
\usepackage{hyperref}% add hypertext capabilities
\usepackage{todonotes}
\usepackage[export]{adjustbox}
\usepackage[ampersand]{easylist}
\usepackage{textcomp}
\usepackage[version=4]{mhchem}
\usepackage{siunitx}
\usepackage[english,printdayoff]{isodate}

\newcommand{\cevns}[1][ ]{CEvNS{#1}}

\newcommand{\cenns[1]}[ ]{CENNS-10{#1}}
\newcommand{\Cs}{\ce{^{137}Cs} }
\newcommand{\Cf}{\ce{^{252}Cf} }
\newcommand{\nue}[1][ ]{\ensuremath{\nu_{e}}{#1}}
\newcommand{\numu}[1][ ]{\ensuremath{\nu_{\mu}}{#1}}

\newcommand{\numubar}[1][ ]{\ensuremath{\overline{\nu}_{\mu}}{#1}}

\newcommand{\fprompt}[1][prompt]{\ensuremath{{F_{{#1}}}} }

\newcommand{\Ar}[1][]{\ce{^{#1}Ar}}

\newcommand{\about}{{\fontfamily{ptm}\selectfont\texttildelow}}

\newcommand{\bigO}[1][]{\ensuremath{\mathcal{O}}{#1}}
\newcommand{\piminus}[1][ ]{\ensuremath{\pi^{-}}{#1}}
\newcommand{\piplus}[1][ ]{\ensuremath{\pi^{+}}{#1}}
\newcommand{\muminus}[1][ ]{\ensuremath{\mu^{-}}{#1}}
\newcommand{\muplus}[1][ ]{\ensuremath{\mu^{+}}{#1}}

\newcommand{\geant}[1][ ]{\texttt{Geant4}{#1}}

\newcommand{\liqar}[1][ ]{\ce{LAr}{#1}}

\newcommand{\fom}[1][ ]{\ensuremath{\mathcal{F}}{#1}}

\newcommand{\csi}[1][ ]{\ce{CsI[Na]}{#1}}

\begin{document}
\title{First Constraint on Coherent Elastic Neutrino-Nucleus Scattering in Argon} 

% downloaded from https://code.ornl.gov/COHERENT/authors/blob/master/authors.tex,  6/17/19
% Updated 7/2 following Yuri's comments
%   Removed LBNL collaborators
%   Y.E. 2nd institution ORNL not MEPhI
%   Should PNNL be removed? And Juan?
% 7/23/19:  Updated by Yuri and passed back to us. Was authors_LAr0.tex
% 8/9/19 Implementing fixes collected from various in response to V2 email, adding newcommand trick so we dont repeat same institutions

\newcommand{\itep}{Institute for Theoretical and Experimental Physics named by A.I. Alikhanov of National Research Centre ``Kurchatov Institute'', Moscow, 117218, Russian Federation}
\newcommand{\mephi}{National Research Nuclear University MEPhI (Moscow Engineering Physics Institute), Moscow, 115409, Russian Federation}
\newcommand{\indiana}{Department of Physics, Indiana University, Bloomington, IN, 47405, USA}
\newcommand{\duke}{Department of Physics, Duke University, Durham, NC 27708, USA}
\newcommand{\tunl}{Triangle Universities Nuclear Laboratory, Durham, NC 27708, USA}
\newcommand{\utk}{Department of Physics and Astronomy, University of Tennessee, Knoxville, TN 37996, USA}
\newcommand{\ornl}{Oak Ridge National Laboratory, Oak Ridge, TN 37831, USA}
\newcommand{\sandia}{Sandia National Laboratories, Livermore, CA 94550, USA}
\newcommand{\fermi}{Enrico Fermi Institute and Kavli Institute for Cosmological Physics, University of Chicago, Chicago, IL 60637, USA}
\newcommand{\chicago}{Department of Physics, University of Chicago, Chicago, IL 60637, USA}
\newcommand{\nmsu}{Department of Physics, New Mexico State University, Las Cruces, NM 88003, USA}
\newcommand{\lanl}{Los Alamos National Laboratory, Los Alamos, NM, USA, 87545, USA}
\newcommand{\cenpa}{Center for Experimental Nuclear Physics and Astrophysics \& Department of Physics, University of Washington, Seattle, WA 98195, USA}
\newcommand{\ncsu}{Department of Physics, North Carolina State University, Raleigh, NC 27695, USA}
\newcommand{\usd}{Physics Department, University of South Dakota, Vermillion, SD 57069, USA}
\newcommand{\virgtech}{Center for Neutrino Physics, Virginia Tech, Blacksburg, VA 24061, USA}
\newcommand{\nccu}{Department of Mathematics and Physics, North Carolina Central University, Durham, NC 27707, USA}
\newcommand{\cmu}{Department of Physics, Carnegie Mellon University, Pittsburgh, PA 15213, USA}
\newcommand{\florida}{Department of Physics, University of Florida, Gainesville, FL 32611, USA}
\newcommand{\laurentian}{Department of Physics, Laurentian University, Sudbury, Ontario P3E 2C6, Canada}
\newcommand{\tufts}{Department of Physics and Astronomy, Tufts University, Medford, MA 02155, USA}
\newcommand{\kaist}{Department of Physics at Korea Advanced Institute of Science and Technology (KAIST)
and Center for Axion and Precision Physics Research (CAPP) at Institute for Basic Science (IBS), Daejeon, 34141, Republic of Korea}

\author{D.~Akimov}\affiliation{\itep}\affiliation{\mephi}
\author{J.B.~Albert}\affiliation{\indiana}
\author{P.~An}\affiliation{\duke}\affiliation{\tunl}
\author{C.~Awe}\affiliation{\duke}\affiliation{\tunl}
\author{P.S.~Barbeau}\affiliation{\duke}\affiliation{\tunl}
\author{B.~Becker}\affiliation{\utk}
\author{V.~Belov}\affiliation{\itep}\affiliation{\mephi}
\author{M.A.~Blackston}\affiliation{\ornl}
\author{A.~Bolozdynya}\affiliation{\mephi}
\author{B.~Cabrera-Palmer}\affiliation{\sandia}
\author{M.~Cervantes}\affiliation{\duke}
\author{J.I.~Collar}\affiliation{\fermi}\affiliation{\chicago}
\author{R.L.~Cooper}\affiliation{\nmsu}\affiliation{\lanl}
\author{J.~Daughhetee}\affiliation{\utk}
\author{M.~del~Valle~Coello}\affiliation{\indiana}
\author{J.A.~Detwiler}\affiliation{\cenpa}
\author{M.~D'Onofrio}\affiliation{\indiana}
\author{Y.~Efremenko}\affiliation{\utk}\affiliation{\ornl}
\author{E.M.~Erkela}\affiliation{\cenpa}
\author{S.R.~Elliott}\affiliation{\lanl}
\author{L.~Fabris}\affiliation{\ornl}
\author{M.~Febbraro}\affiliation{\ornl}
\author{W.~Fox}\affiliation{\indiana}
\author{A.~Galindo-Uribarri}\affiliation{\utk}\affiliation{\ornl}
\author{M.P.~Green}\affiliation{\tunl}\affiliation{\ornl}\affiliation{\ncsu}
\author{K.S.~Hansen}\affiliation{\cenpa}
\author{M.R.~Heath}\affiliation{\indiana}
\author{S.~Hedges}\affiliation{\duke}\affiliation{\tunl}
\author{T.~Johnson}\affiliation{\duke}\affiliation{\tunl}
\author{M.~Kaemingk}\affiliation{\nmsu}
\author{L.J.~Kaufman\footnote{Also at SLAC National Accelerator Laboratory, Menlo Park, CA 94205, USA}}\affiliation{\indiana}
\author{A.~Khromov}\affiliation{\mephi}
\author{A.~Konovalov}\affiliation{\itep}\affiliation{\mephi}
\author{E.~Kozlova}\affiliation{\itep}\affiliation{\mephi}
\author{A.~Kumpan}\affiliation{\mephi}
\author{L.~Li}\affiliation{\duke}\affiliation{\tunl}
\author{J.T.~Librande}\affiliation{\cenpa}
\author{J.M.~Link}\affiliation{\virgtech}
\author{J.~Liu}\affiliation{\usd}
\author{K.~Mann}\affiliation{\tunl}\affiliation{\ornl}
\author{D.M.~Markoff}\affiliation{\tunl}\affiliation{\nccu}
\author{H.~Moreno}\affiliation{\nmsu}
\author{P.E.~Mueller}\affiliation{\ornl}
\author{J.~Newby}\affiliation{\ornl}
\author{D.S.~Parno}\affiliation{\cmu}
\author{S.~Penttila}\affiliation{\ornl}
\author{D.~Pershey}\affiliation{\duke}
\author{D.~Radford}\affiliation{\ornl}
\author{R.~Rapp}\affiliation{\cmu}
\author{H.~Ray}\affiliation{\florida}
\author{J.~Raybern}\affiliation{\duke}
\author{O.~Razuvaeva}\affiliation{\itep}\affiliation{\mephi}
\author{D.~Reyna}\affiliation{\sandia}
\author{G.C.~Rich}\affiliation{\fermi}
\author{D.~Rudik}\affiliation{\itep}\affiliation{\mephi}
\author{J.~Runge}\affiliation{\duke}\affiliation{\tunl}
\author{D.J.~Salvat}\affiliation{\indiana}\affiliation{\cenpa}
\author{K.~Scholberg}\affiliation{\duke}
\author{A.~Shakirov}\affiliation{\mephi}
\author{G.~Simakov}\affiliation{\itep}\affiliation{\mephi}
\author{G.~Sinev}\affiliation{\duke}
\author{W.M.~Snow}\affiliation{\indiana}
\author{V.~Sosnovtsev}\affiliation{\mephi}
\author{B.~Suh}\affiliation{\indiana}
\author{R.~Tayloe}\affiliation{\indiana}
\author{K.~Tellez-Giron-Flores}\affiliation{\virgtech}
\author{R.T.~Thornton}\affiliation{\indiana}\affiliation{\lanl}
\author{I.~Tolstukhin}\affiliation{\indiana}
\author{J.~Vanderwerp}\affiliation{\indiana}
\author{R.L.~Varner}\affiliation{\ornl}
\author{C.J.~Virtue}\affiliation{\laurentian}
\author{G.~Visser}\affiliation{\indiana}
\author{C.~Wiseman}\affiliation{\cenpa}
\author{T.~Wongjirad}\affiliation{\tufts}
\author{J.~Yang}\affiliation{\tufts}
\author{Y.-R.~Yen}\affiliation{\cmu}
\author{J.~Yoo}\affiliation{\kaist}
\author{C.-H.~Yu}\affiliation{\ornl}
\author{J.~Zettlemoyer}\affiliation{\indiana}\affiliation{\ornl}
  
\date{September 2019}

\begin{abstract}
Coherent elastic neutrino-nucleus scattering (\cevns[]) is the dominant neutrino scattering channel for neutrinos of energy $E_\nu<$\SI{100}{\MeV}. We report a limit for this process using data collected in an engineering run of the \SI{29}{kg} \cenns liquid argon detector located \SI{27.5}{m} from the Oak Ridge National Laboratory Spallation Neutron Source (SNS) Hg target with $4.2\times 10^{22}$ protons on target. The dataset yielded \textless\num{7.4} observed \cevns events implying a cross section for the process, averaged over the SNS pion decay-at-rest flux, of \textless\SI{3.4e-39}{\cm\squared}, a limit within twice the Standard Model prediction. This is the first limit on \cevns from an argon nucleus and confirms the earlier \csi non-standard neutrino interaction constraints from the collaboration. This run demonstrated the feasibility of the ongoing experimental effort to detect \cevns with liquid argon.
\end{abstract}
\maketitle

\section{\label{sec:intro}Introduction}
Coherent elastic neutrino-nucleus scattering (\cevns[]), predicted in 1974 as a consequence of the neutral weak current~\cite{Freedman:1973yd,Kopeliovich:1974mv}, is the dominant neutrino interaction for neutrinos of energy $E_\nu < \SI{100}{\MeV}$. 
It has a characteristic dependence on the square of the number of neutrons ($N^2$) 
reflecting the coherent sum of the weak charge carried by the neutrons, and is sensitive to  nuclear physics effects~\cite{bib:formfactorscevns, bib:cadedduNeutronRad, bib:cadeddu, Patton:2012jr, AristizabalSierra:2019zmy, Hoferichter:2018acd} through the nuclear form factor, ($F(Q^{2})$), as seen in the differential cross section for a spin-zero nucleus~\cite{bib:formfactorscevns}:
\begin{equation} \label{eq:xsec}
\frac{d\sigma}{dT} = \frac{G_{F}^{2}M}{2\pi}\left[ 2 - \frac{2T}{E_{\nu}} + \left( \frac{T}{E_{\nu}}\right)^{2} -\frac{MT}{E_{\nu}^{2}}\right]\frac{Q_{W}^{2}}{4}F^{2}(Q^{2})
\end{equation}
where $T$ is the recoil energy, $M$ is the mass of the nucleus, and $Q_{W} = N - Z \left( 1-4\sin^{2}\theta_{W} \right)$ is the weak charge with weak mixing angle $\theta_{W}$.
\cevns is also sensitive to physics beyond the Standard Model (SM)~\cite{bib:barrancoCrossSection, bib:scholberg, bib:barrancoNSI, bib:duttaBSM, bib:papoulias, bib:kraussWeakCharge}.
In particular, the ability of a \cevns measurement to constrain so-called ``Non-Standard Interactions" (NSI) is critical as their presence can confound the mass ordering determination by long-baseline neutrino experiments such as DUNE~\cite{bib:coloma, bib:coloma2, bib:coloma3}.

\cevns has eluded detection until recently due to the challenging technical requirements: $\bigO(\SI{10}{\keV})$ nuclear recoil energy thresholds, intense sources/large target masses, and low backgrounds. The COHERENT collaboration has recently overcome these challenges with state-of-the-art detector technology combined with the intense, pulsed, stopped-pion neutrino source available at the Spallation Neutron Source (SNS) at Oak Ridge National Laboratory (ORNL), using a \csi crystal to achieve the first measurement of \cevns~\cite{Akimov:2017ade}.

The next step for this program is a demonstration of the $N^2$ cross section dependence via observation of the process in other nuclei.
To that end, the \SI{29}{\kg} liquid argon detector \cenns was commissioned as part of the COHERENT experiment.
We report here results from \cenns as configured for an initial engineering run to establish the scintillation response, light yield, and energy calibration of the detector, as well as characterize the expected backgrounds. The results reported here informed a detector upgrade for a longer-term \cevns search with improved light yield and background reduction.

\begin{figure}
\includegraphics[width=0.99\columnwidth,trim={85 60 55 50},clip]{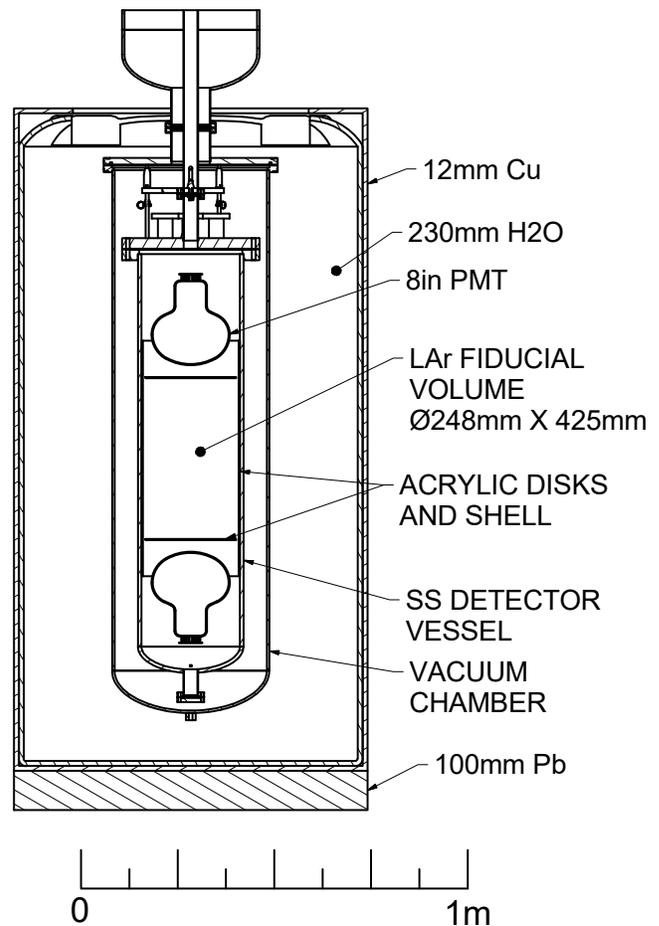}
\caption{\label{fig:CENNS-10} \cenns liquid argon detector and shielding as configured for this engineering run.}
\end{figure}

\section{\label{sec:exp} Experiment}
The ORNL SNS produces neutrons via a \SI{1.4}{\MW}, \SI{1}{\GeV} proton beam pulsed at \SI{60}{\Hz} on a liquid-\ce{Hg} target (with a typical proton beam trace having a $\mathrm{FWHM} = \SI{360}{\ns}$). This beam also produces copious charged pions leading to a large neutrino flux via \piplus decay-at-rest (DAR). 
While the total integrated beam power may be known to \SI{<1}{\percent}, the total neutrino flux is only known to \SI{10}{\percent} due to systematic uncertainties in the pion production rate at the SNS, predicted to be \SI{0.09}{\piplus} for each proton-on-target (POT) at the beam energy for this run period~\cite{Akimov:2017ade}.
These \piplus produce a prompt ($\tau = \SI{26}{\ns}$) \SI{29.8}{\MeV} \numu along with a \muplus which then decays, yielding a delayed ($\tau=\SI{2.2}{\micro\s}$) 3-body spectrum of \numubar[], \nue[], and $e^{+}$ with an endpoint of \SI{53}{\MeV}. The majority of \piminus and \muminus capture on nuclei within the target yielding a very pure \piplus DAR neutrino flux. The pulsed nature of the SNS beam allows for a large reduction in beam-unrelated backgrounds for neutrino experiments.

After a campaign of background measurements in the SNS experimental hall, a low-background area in a basement corridor was identified as a suitable area in which to measure \cevns[]. This corridor (``Neutrino Alley''), is shielded by $\gtrsim\SI{20}{\m}$ of concrete and gravel from the SNS target assembly and by \SI{8}{meter~water~equivalent} overburden. This provides a space with a low total background rate and, in particular, a sufficiently low beam-related-neutron rate for a measurement of \cevns[]. 

In late 2016 the \cenns detector, a single-phase liquid-argon (\liqar[]) scintillation detector (Fig.~\ref{fig:CENNS-10})~\cite{Tayloe:2017edz}, 
was installed in Neutrino Alley \SI{27.5}{\m} from the SNS target. \cenns was initially built at Fermilab for a prototype experiment~\cite{bib:CENNS} to run near the Fermilab Booster neutrino source. It contains a total \liqar mass of \SI{79.5}{\kg}.

For this engineering run, a \SI{29}{\kilo\gram} active detector mass was defined by a surrounding  acrylic cylindrical shell coated with \SI{0.2}{\mg\per\cm\squared} TPB (tetraphenyl-butadiene) to wavelength-shift the \SI{128}{\nm} argon scintillation light to a distribution with $\lambda_{peak} \approx \SI{400}{\nm}$~\cite{bib:tpbeff1, bib:tpbeff2, bib:mckinsey, bib:agnesTPB}. This visible light was viewed with two 8" diameter Hamamatsu R5912-02MOD photomultiplier tubes (PMTs) read out with a CAEN V1720 digitizer. The \liqar[], cooled and liquified with a \SI{90}{\W} Cryomech PT-90 cold head, was contained in a stainless-steel detector vessel within a vacuum cryostat. As seen in Fig.~\ref{fig:CENNS-10}, the cryostat was suspended in a cylindrical water tank  which was further contained within an external copper layer sitting on a layer of lead. The water layer reduces the beam-related neutron backgrounds, the lead is designed to reduce the flux from environmental $\gamma$ backgrounds, and the copper is added to shield from x-rays produced from \ce{{}^{210}Pb} $\beta$ decays in the lead. 

\begin{figure}
\centerline{\includegraphics[width=0.99\columnwidth]{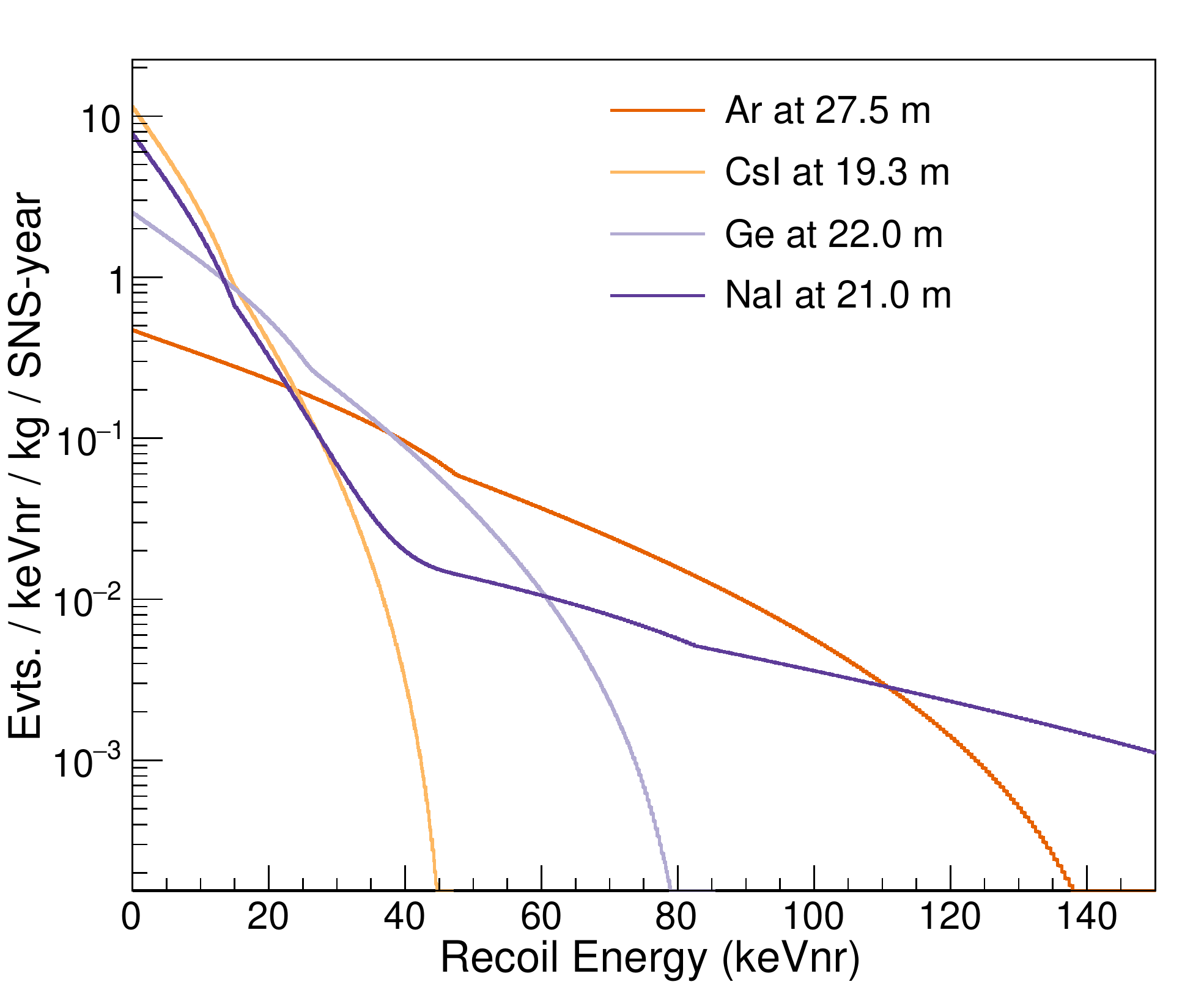}}
\caption{\label{fig:CEvNSrecoilE} Nuclear recoil kinetic energy distribution from \cevns for the SNS neutrino spectrum for currently-deployed and planned COHERENT detectors at their respective detector locations in Neutrino Alley. 
}
\end{figure}

This engineering run coincided with three months of SNS neutron production corresponding to a total integrated beam power of \SI{1.8}{\GW hr} (\SI{4.2e22}{POT}) at an average energy of \SI{973}{MeV}. A \cevns search was performed with \SI{1.5}{\GW hr} of beam following the completion of the full-shielding (water and copper) installation. Data were read from the digitizer in \SI{33}{\micro\s} windows centered around the \SI{60}{\Hz} beam spills. In addition to these ``beam'' triggers, identical windows (``strobe'' triggers) were read asynchronously with the beam spills to precisely characterize beam-unrelated events. 

\liqar is a natural choice as a medium to detect \cevns[]. It provides a light nucleus in contrast to CsI to test the $N^{2}$ dependence of the \cevns cross section. Argon has been widely used for both dark matter WIMP searches~\cite{PhysRevD.98.102006,Ajaj:2019imk} and for neutrino detection~\cite{PhysRevD.99.091102}, and has therefore been well-characterized in the literature. It has a high light yield, \SI{40}{photons\per\keV ee}~\cite{bib:Doke90} (electron equivalent energy deposition), providing a sufficiently low threshold for \cevns detection, and the quenched response to nuclear recoils has been well-characterized~\cite{bib:microclean, bib:scene, bib:creus, bib:aris} allowing for well-understood \cevns predictions.
\liqar scintillates on two significantly different time scales ($\tau_{singlet} \approx \SI{6}{\ns}, \tau_{triplet} \approx \SI{1600}{\ns}$)~\cite{bib:Hitachi83} providing powerful pulse-shape discrimination (PSD) capabilities to separate nuclear from electronic recoils (NR and ER respectively)~\cite{bib:deap1, bib:benetti, bib:lippincott}. Both the light output and PSD capabilities depend on the \liqar purity.

As seen in Fig.~\ref{fig:CEvNSrecoilE}, the \cevns process in \liqar with the SNS neutrino source produces nuclear recoils up to \about\SI{100}{\keV nr} (nuclear recoil). Due to the low-energy recoil signal, and the low event rates, the expected backgrounds need to be well characterized. In Neutrino Alley, \cenns is sensitive to both beam-related and beam-unrelated backgrounds. These beam-unrelated backgrounds typically cause electronic recoils and are dominated by a high flux of \SI{511}{\keV} gamma rays from a pipe running through Neutrino Alley carrying radioactive gas from the SNS target system. The PSD capabilities of \liqar are used to reject most of these events; the rate of those remaining in the sample is measured via the strobe windows. In a strict sense, these \SI{511}{\keV} gamma rays are beam-related and their rates change with the time history of accelerator operations. However, as the rate of change is small compared to the beam pulse rate, they are characterized as beam-unrelated.  
External beam-unrelated backgrounds have largely been mitigated in a subsequent run of \cenns with the installation of additional \ce{Pb} shielding, making \Ar[39] the dominant beam-unrelated background. The \Ar[39] isotope is cosmogenically produced and is inherent in atmospheric sources of \ce{Ar}.
COHERENT is considering the use of underground argon depleted in \Ar[39]~\cite{bib:backuar, bib:backuar2, bib:agnesuar} for future \liqar measurements.

A more challenging background for a \cevns analysis is caused by beam-related neutrons (BRNs) produced in the SNS target. BRNs arrive in-time with the SNS beam pulse and elastically scatter, generating nuclear recoils and mimicking the \cevns signal. To characterize the BRN flux in energy and time, it was measured by the SciBath detector~\cite{Cooper:2011kx,bib:CENNS} at the \cenns location in late 2015. This measurement indicated that the BRN flux in time with the beam pulse is substantial compared to the prompt \cevns signal while the delayed BRN flux is negligible, thus providing a suitable time window in which to search for \cevns[]~\cite{Heath:2019jpj}.

\section{\label{sec:anl} Analysis}
The analysis of this dataset proceeded as follows: First a suite of radioactive $\gamma$ and neutron sources were used to calibrate the detector energy and PSD response and the detector simulation was tuned to match these data.
Then beam-unrelated backgrounds were measured with strobe triggers, the beam-related background from BRNs was predicted with simulation based on the previous SciBath measurement, and the \cevns signal was predicted from the SM cross section.  
Energy, PSD, and time cuts were then optimized with those estimates to maximize beam-related signal significance. With those cuts, a reduced neutron-shielding dataset was used to adjust the BRN prediction for the full shielded configuration.  
Finally, cuts were optimized and fixed for both a `counting experiment' and a likelihood fit before analyzing the full shielded beam-on dataset. 

The individual, digitized PMT waveforms are analyzed for every trigger in the data stream and saturated waveforms are removed from the dataset. A baseline is determined from the average ADC value in the first \SI{1}{\micro\s} of each remaining waveform. This baseline is then used to identify PMT pulses on each channel above a \SI{20}{ADC} (\about\SI{2}{photoelectron}) threshold. Events are identified when there are coincident PMT signals above this threshold to avoid triggering on single photoelectron-level pulses from PMT dark rate. A requirement that the maximum ADC value occur within the first \SI{90}{\ns} of the event minimizes the effects from event pileup. A local baseline is calculated immediately before each pulse and a least-squares parabola fit is performed to the pulse peak for an accurate singlet pulse-height measurement. The results from the parabola fit are used to fit a single photoelecton (SPE) template shape to the singlet peak and the residual between the SPE template and the data is taken. Finally, the integral of the residual waveform is taken as a measure of the amount of triplet light in the event. A pulse shape parameter (\fprompt defined as the ratio of singlet to total light) can then be calculated to separate ER background events from the NR \cevns signal. 
\begin{figure}
    \centerline{\includegraphics[width=0.99\columnwidth]{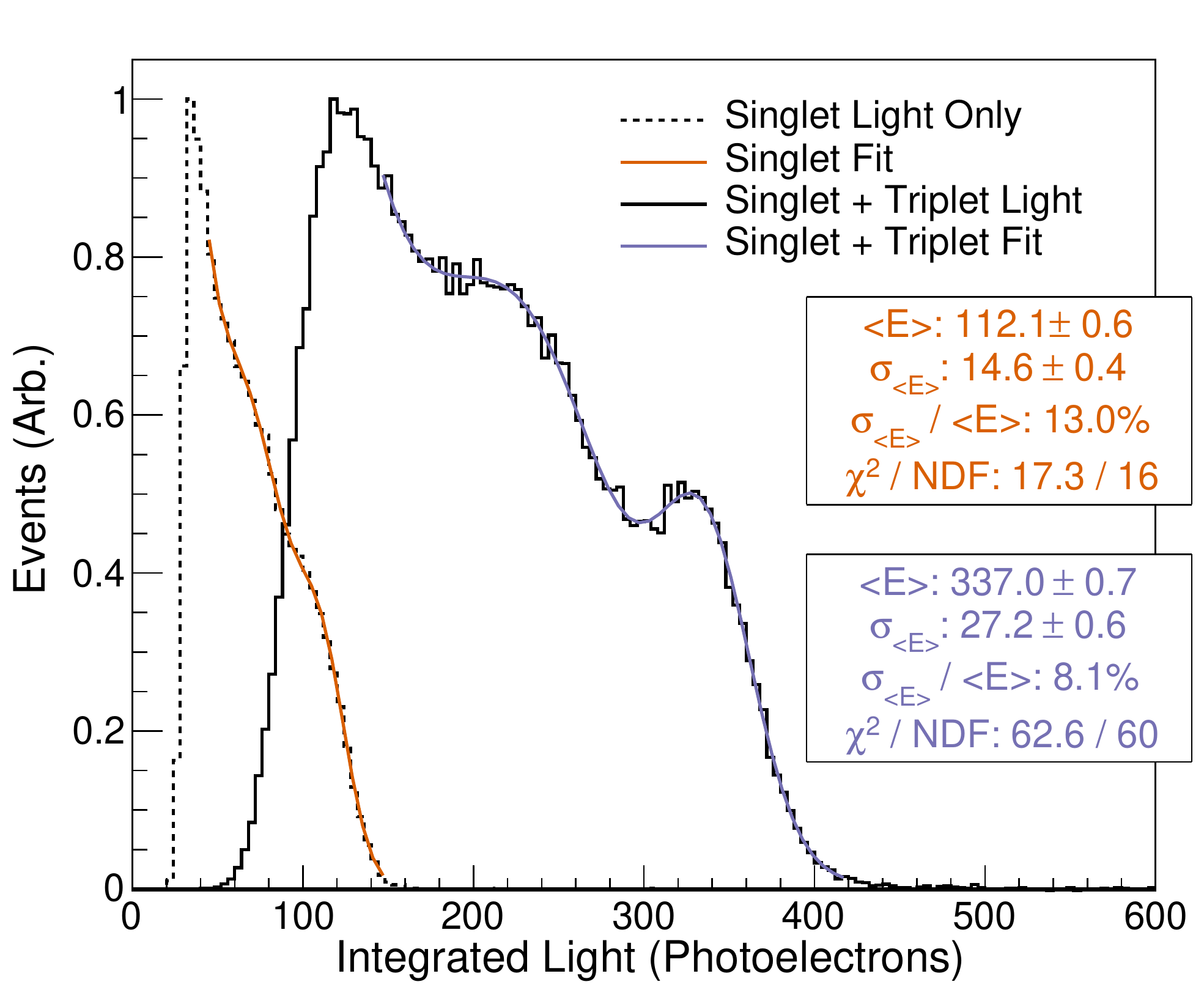}}
    \caption{Reconstructed energy spectrum with a \Cs source. The singlet pulse fitting allows for the singlet light to be summed separately.}
    \label{fig:ly}
\end{figure}

Weekly calibration datasets with a \Cs source were used to measure the detector light output as well as track any changes over the course of this run. The detected photon yield was \SI{0.6}{PE\per\keV ee} as determined from the observed \SI{662}{keV} photopeak from the summed singlet and triplet light in the \Cs spectrum (Fig.~\ref{fig:ly}).  It should be noted that the light yield was increased by a factor of \num{8} in a subsequent upgrade of this detector. With the use of the \Cs datasets, the triplet lifetime in \cenns was measured to be $\bigO(\SI{1.2}{\micro\s})$, consistent with an impurity level on the order of $\bigO(\SI{1}{ppm})$~\cite{bib:WArP10}, adequate for a scintillation-only detector.

\begin{figure}
    \centerline{\includegraphics[width=0.99\columnwidth]{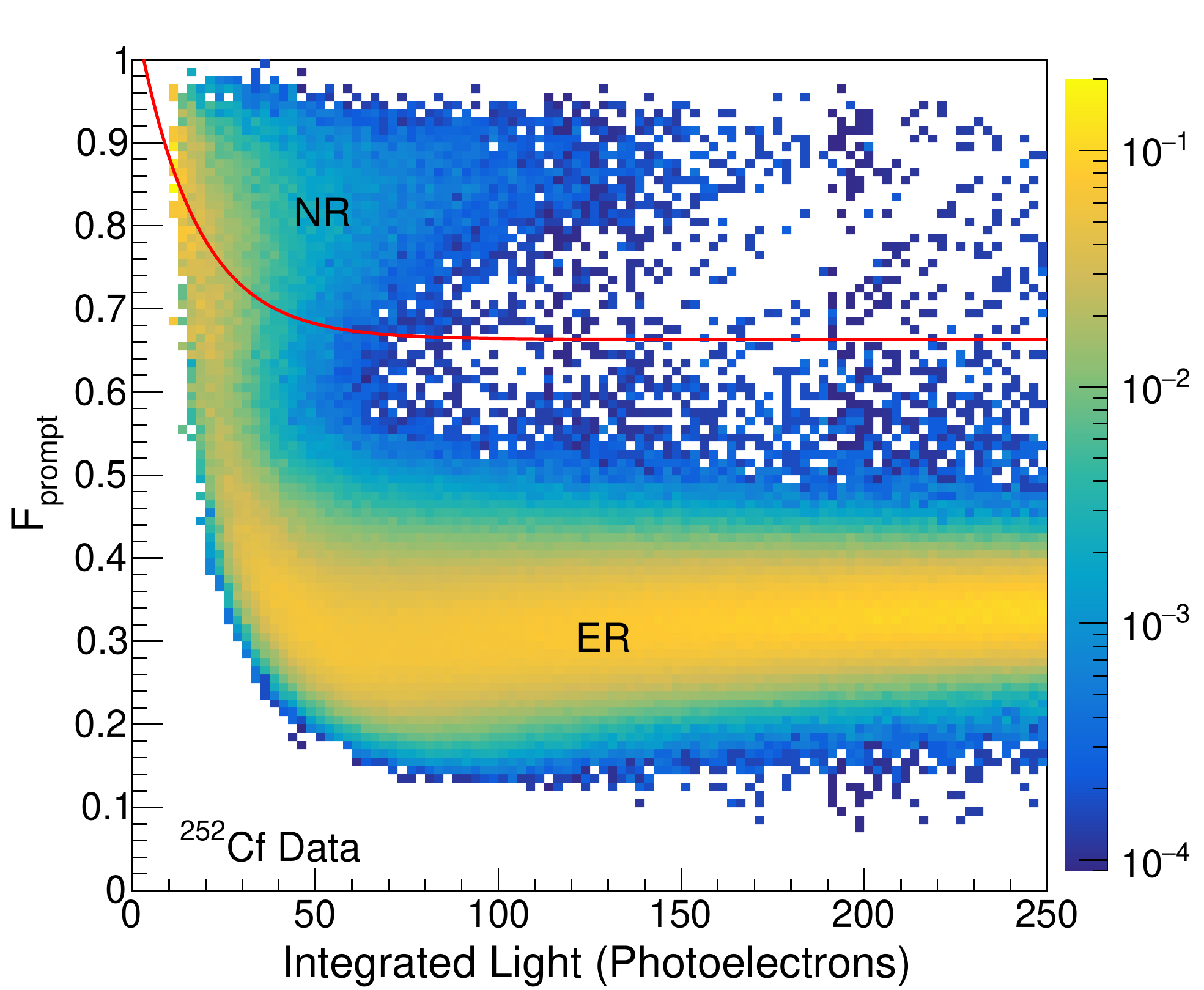}}
    \caption{Distribution of the \fprompt parameter as a function of detected light in \Cf calibration data with decays to both neutrons and $\gamma$s. The overlaid red curve is a PSD cut optimized for the cuts-based counting experiment analysis discussed in the text.}
    \label{fig:arpsd}
\end{figure}

Monthly datasets  collected with a \Cf fission source were used to characterize the response of \cenns to NR events. 
The separation of NR and ER events in the \Cf dataset is shown in Fig.~\ref{fig:arpsd} where the band at low \fprompt is identified as due to ER events and that at high \fprompt is identified as NR events due to the fission neutrons. The observed \fprompt is consistent with the expected singlet:triplet ratios of ER and NR events~\cite{bib:Hitachi83}.

These calibration datasets enabled the tuning of the \cenns \geant[]-based~\cite{Agostinelli:2002hh} Monte Carlo (MC) simulation optical properties for both ER and NR events. 
These detector simulations were used to evaluate the efficiency for low-energy NR events to be detected and to form predictions of the expected BRN and \cevns event rates in \cenns[]. An energy-independent fit over the energy range of interest  to the global \liqar data on nuclear recoil scintillation quenching~\cite{bib:microclean, bib:scene, bib:creus, bib:aris} provided a quenching factor ($0.289 \pm 0.035$) for NR vs ER response in \cenns[]. With these waveform analysis and calibration procedures, each detector event can be identified as an ER or NR candidate and be assigned a corresponding energy with units of \si{\keV ee} or \si{\keV nr}. 

Initial BRN predictions using a simulation based on the 2015 SciBath measurement were compared to a dedicated two-week minimal-neutron-shielding dataset. 
From this comparison, the predicted BRN rate was found to be \SI{20}{\percent} lower than the observed rate.
This factor was used to adjust the expected neutron rates for the primary \cevns dataset. However, the BRN normalization was allowed to float in the final analysis.
\cevns predictions were based on the convolution of the pion decay-at-rest neutrino flux and SNS pion-production rate~\cite{Akimov:2017ade} with the Standard Model-predicted \cevns cross section. Beam-unrelated backgrounds were measured \textit{in situ} with strobe triggers.

Both a cuts-based (``counting experiment'') analysis and a likelihood fit in energy, time, and \fprompt space were performed on the full-shielded \cevns dataset.  In the cuts-based analysis, to form a \cevns\ sample, a figure-of-merit $\fom \equiv N_{sig} / \sigma_{sig}$ was optimized to set a \SIrange{0}{30}{\keV ee} reconstructed energy range, a delayed $1.4 < t_{Trig} < \SI{4.4}{\micro\s}$ time window (where $t_{Trig}$ is measured relative to a timing signal provided by the SNS close to the onset of POT), and an energy-dependent PSD selection seen in Fig.~\ref{fig:arpsd}. For this analysis, it was assumed that the BRNs observed in Neutrino Alley are produced by fast neutrons from the target scattering in the shielding near the detector and that the neutrinos should arrive roughly \SI{30}{\ns} before the fast neutron peak determined from the BRN measurements.  The results reported here are not sensitive to this assumption.  
A BRN-enhanced sample was selected with an expanded energy range (\SIrange{0}{700}{\keV ee}) in both the prompt ($0.4 < t_{Trig} < \SI{1.4}{\micro\s}$) and the delayed ($1.4 < t_{Trig} < \SI{4.4}{\micro\s}$) time windows.
\begin{figure}
    \centerline{\includegraphics[width=0.99\columnwidth]{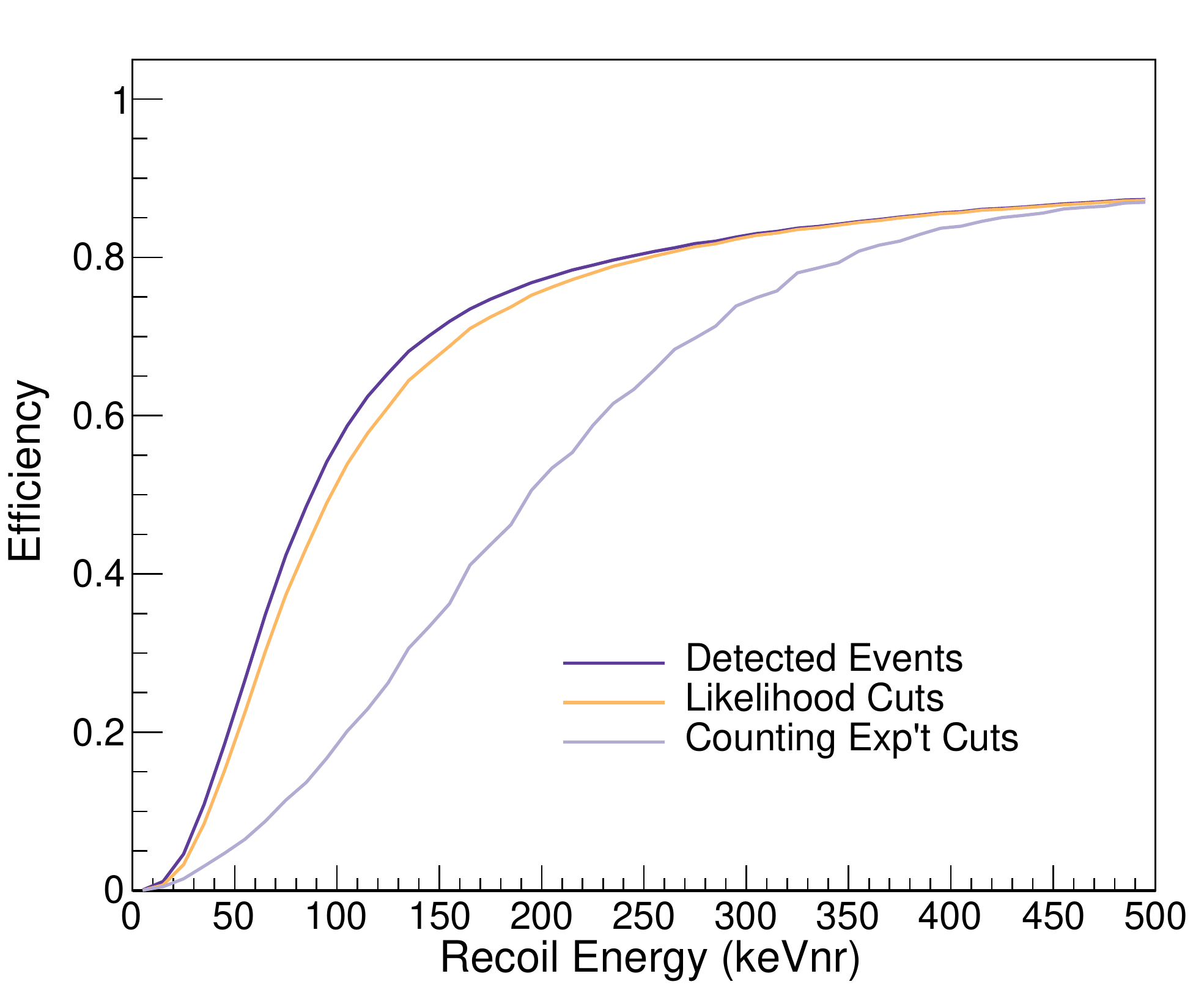}}
    \caption{Estimated efficiency for acceptance of nuclear recoil events in \cenns as function of nuclear recoil energy. ``Detected Events'' are those that pass the \SI{2}{PE} coincidence required for event building. The likelihood and counting experiment cuts reflect the change in efficiency due to analysis cuts discussed in the text.}
    \label{fig:eff}
\end{figure}

For the likelihood fit, cuts were loosened, increasing the sensitivity to a \cevns signal, to \SIrange{0}{300}{\keV ee}, \SIrange{0.4}{4.4}{\micro\s} relative to the SNS timing signal, and from \fprompt values ranging from \SIrange{0.55}{0.95}{}. The lack of \cevns events with reconstructed energy $E_{reco} > \SI{50}{\keV ee}$ and the lack of BRN events in the delayed window ($t_{Trig} > \SI{1.4}{\micro\s}$) serves to separate the BRN and \cevns signals. The efficiencies as a function of nuclear recoil energy for these cuts is seen in Fig.~\ref{fig:eff}.

Systematic errors were assigned to the beam-related (\cevns and BRN) predictions for the quenching factor and pulse-finding threshold. These uncertainties were dominated by the uncertainty of the NR PSD band in the \cevns energy region due to the high threshold of the \Cf calibration datasets. An additional source of uncertainty was included on the overall BRN normalization due to the extrapolation of the BRN rate from the minimal-shielded dataset. For the cuts-based analysis, correlated systematic errors were calculated and a goodness-of-fit ($\chi^{2}$) quantity was determined for the beam excess compared to the MC prediction.  For the cross section limits from the likelihood fits, alternative PDFs incorporating \SI{\pm1}{\sigma} excursions for each systematic were fit to the data, and the difference from the central value result were added in quadrature as a measure of the systematic uncertainty.

\section{Results} 
\begin{figure}
    \centerline{\includegraphics[width=0.99\columnwidth]{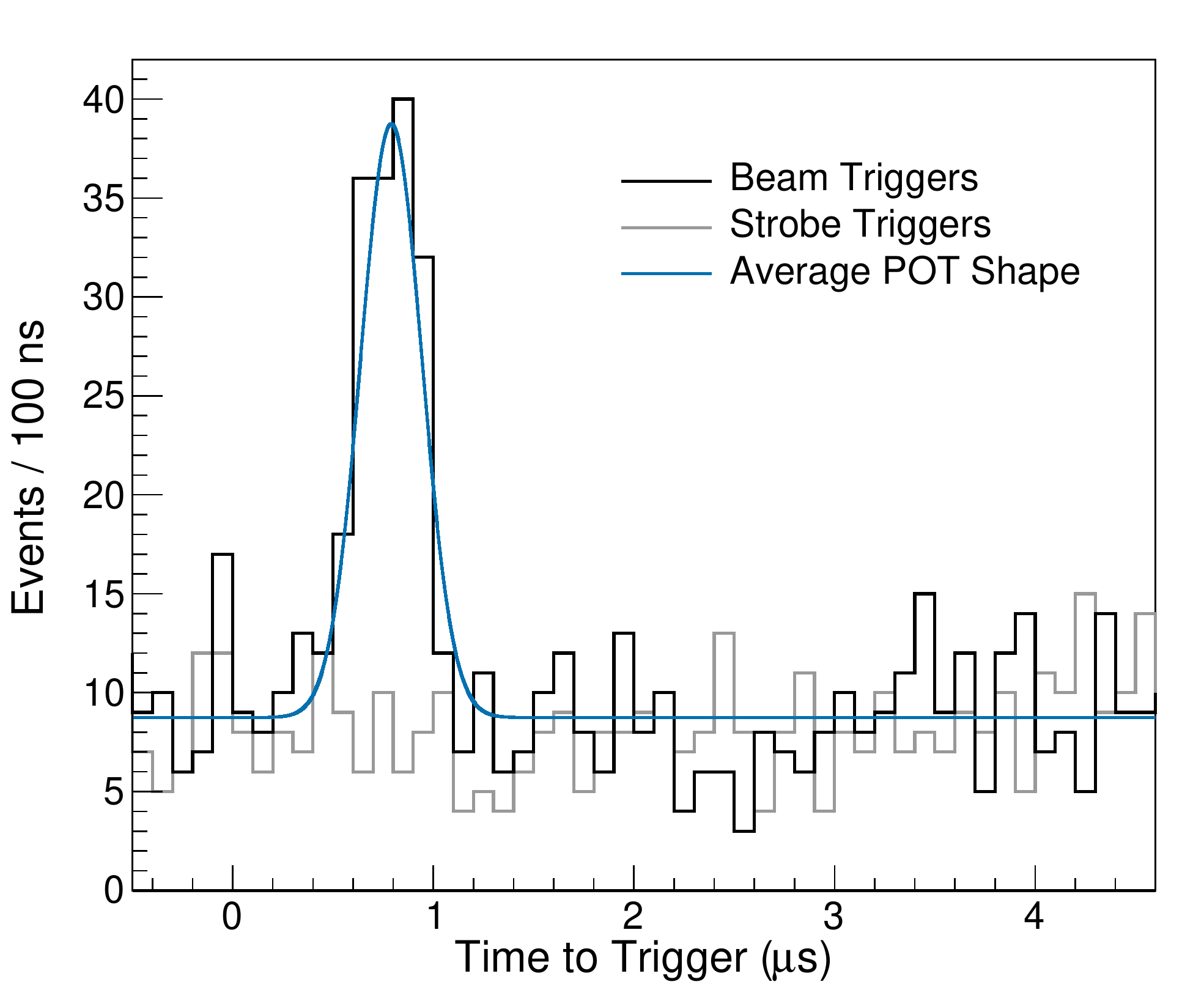}}
    \caption{Time distribution of beam-on and strobe samples in the BRN-enhanced energy window. The blue curve is that expected from the timing shape of the SNS POT signal scaled to the beam-on-target excess.}
    \label{fig:counting_time}
\end{figure}
The resulting sample from the BRN-enhanced cuts-based analysis (\SIrange{0}{700}{\keV ee}) over the full time range is shown in Fig.~\ref{fig:counting_time}.  Note the clear evidence of BRNs with time structure consistent with the POT trace from the SNS beam. Note also that there is no evidence of this signal in the delayed ($t_{Trig} > \SI{1.4}{\micro\s}$) region. This is consistent with the hypothesis that the BRN that reach the \cenns detector inside of the shielding are the result of fast neutrons in Neutrino Alley that lose sufficient energy to create low-energy nuclear recoils in \liqar[]. This is verified by MC simulations. 

\begin{figure}
      \centerline{\includegraphics[width=0.99\columnwidth]{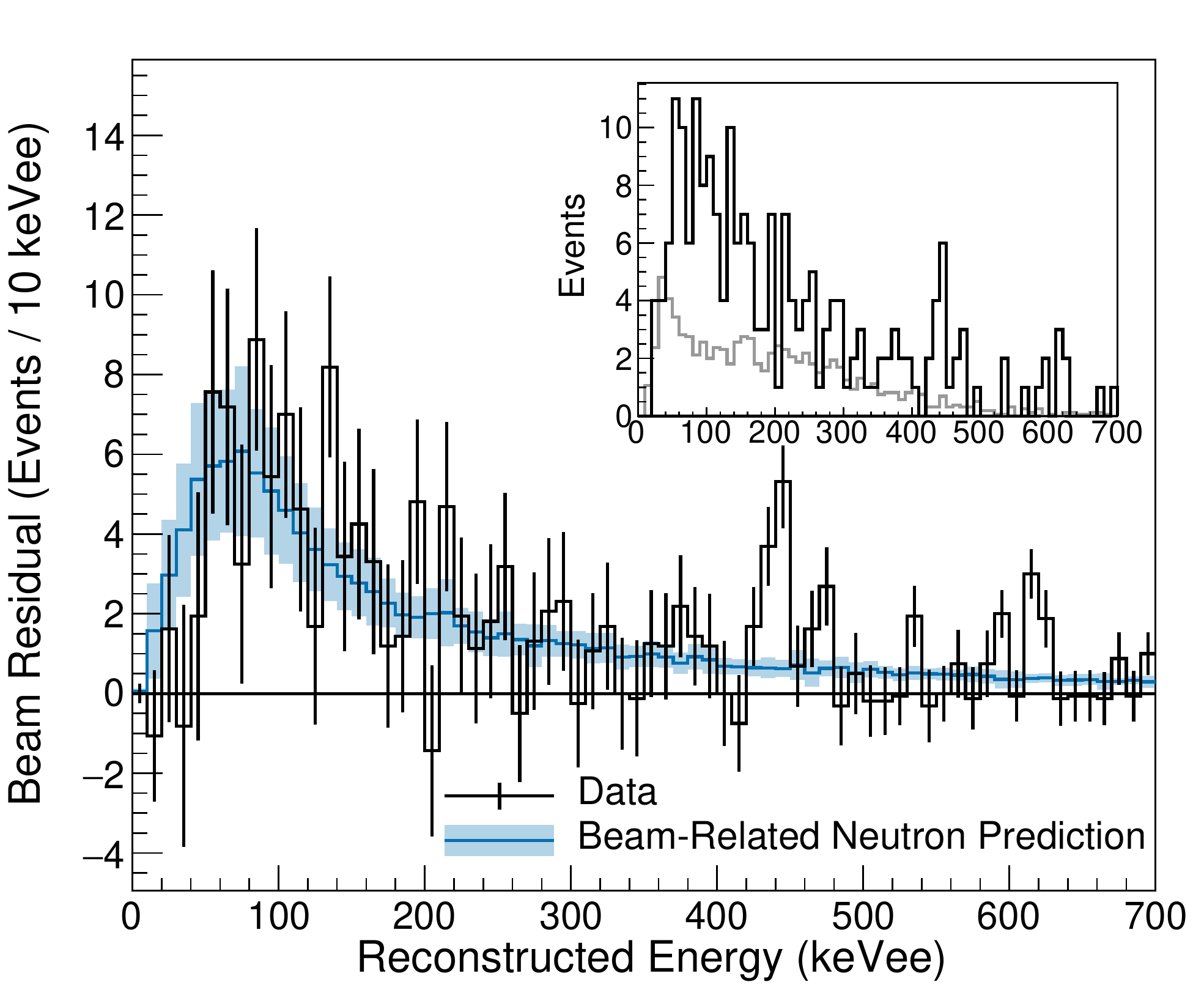}}
    \caption{Energy distribution of the cuts-based analysis beam-residual event sample in the prompt time window along with the BRN prediction. The error bars are statistical and the error band on the prediction is systematic.  Plot inlay shows un-subtracted spectra from the prompt beam-on triggers (black) and the expected beam-unrelated background as measured with strobe triggers (gray). 
    }
    \label{fig:counting_prompt}
\end{figure}

The reconstructed energy distribution from this sample in the prompt time region ($0.4 < t_{Trig} < \SI{1.4}{\micro\s}$) is shown in Fig.~\ref{fig:counting_prompt}. The beam-related excess of \SI[parse-numbers=false]{126 \pm 15(stat.)}{events} in this sample is consistent with the BRN prediction of \SI[parse-numbers=false]{112 \pm 30(syst.)}{events}. 
The uncertainty on the BRN prediction is dominated by the uncertainty in the overall fast neutron flux (\SI{\pm20}{\percent}), the uncertainty of the NR PSD band mean near threshold (\SI{\pm19}{\percent}), the pulse-finding threshold (\SI{\pm5}{\percent}), and the quenching factor (\SI{\pm4}{\percent}). 
The predicted \cevns signal in this sample is \num{<1} detected event.  
A comparison of the data with the predicted BRN energy spectrum gives a $\chi^{2} / N_{bins}$, including correlated uncertainties, of $99 / 70$ 
($2.0 / 3$ in the \cevns energy ROI). The excess of events above prediction at $E \approx \SI{440}{\keV ee}$ has a global p-value under the null hypothesis of \SI{1.7}{\percent} and is above the energy region of interest for the likelihood fit.

\begin{figure}
      \centerline{\includegraphics[width=0.99\columnwidth]{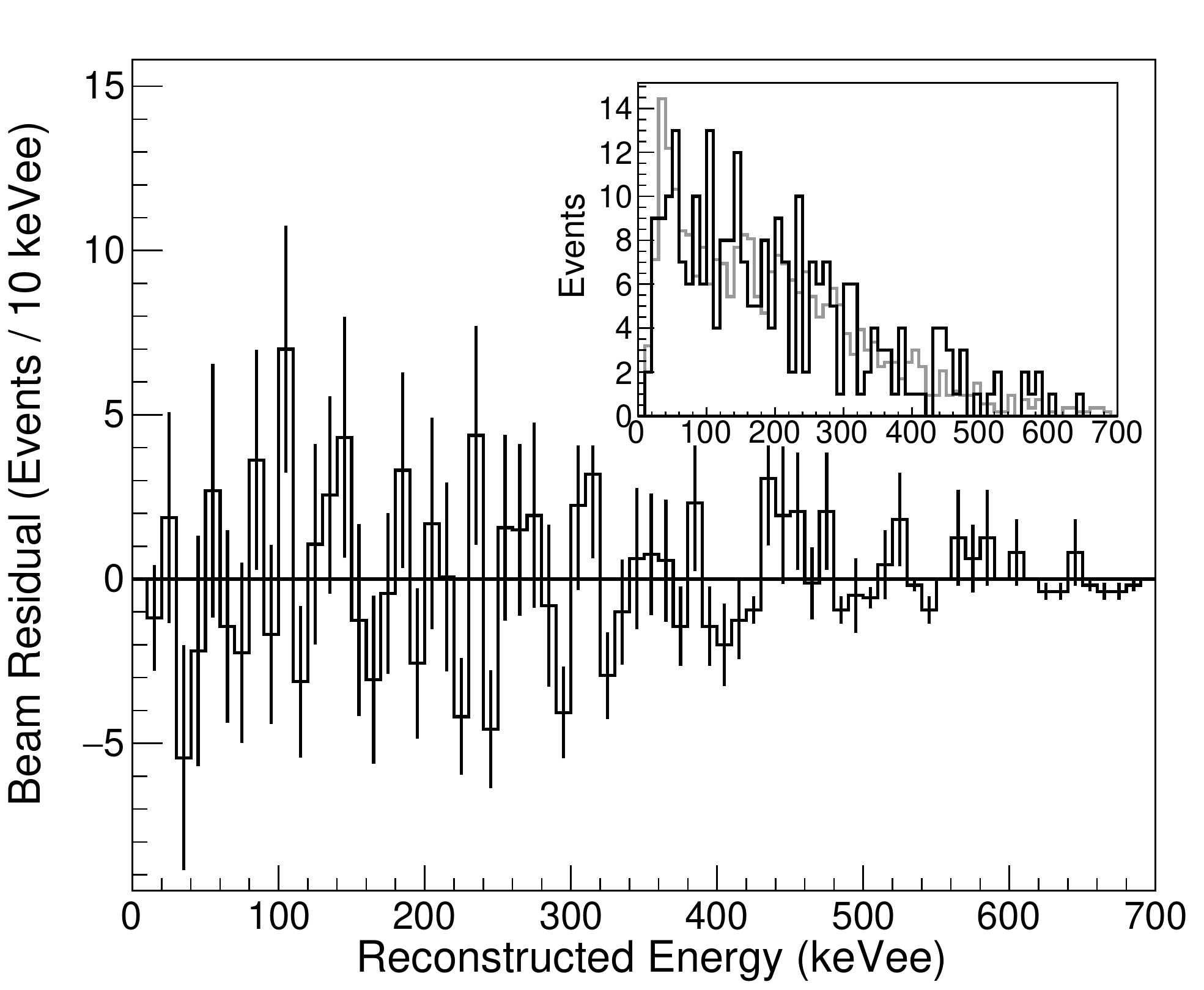}}
      \caption{Energy distribution of the cuts-based analysis beam-residual event sample in the delayed time window.  Plot inlay shows un-subtracted beam-on spectrum (black) along with the expected beam-unrelated backgrounds as measured with strobe triggers (gray).}
      \label{fig:fullpsddelayed}
\end{figure}

The energy distribution of events in the delayed sample is shown in Fig.~\ref{fig:fullpsddelayed}. In the \cevns energy region \SIrange{0}{30}{\keV ee}, an excess of 
\SI[parse-numbers=false]{1 \pm 4(stat.)}{events} is observed, with a predicted \cevns sample of 
\SI{<1}{event} with an uncertainty dominated by the pulse-finding threshold (\SI{\pm35}{\percent}), the NR PSD band mean behavior near threshold (\SI{\pm30}{\percent}), the quenching factor (\SI{\pm15}{\percent}), and the uncertainty in the neutrino flux (\SI{\pm10}{\percent}). The first two errors are large because the CEvNS events are so near the threshold in this dataset.  In addition, there are \SI[parse-numbers=false]{9 \pm 18(stat.)}{events} in the extended energy range out to \SI{700}{\keV ee}, consistent with earlier measurements~\cite{Akimov:2017ade, bib:coherent2015} indicating no delayed beam-related neutron flux in Neutrino Alley.

\begin{figure*}
    \begin{minipage}{0.325\textwidth}
        \includegraphics[width=\textwidth,page=1]{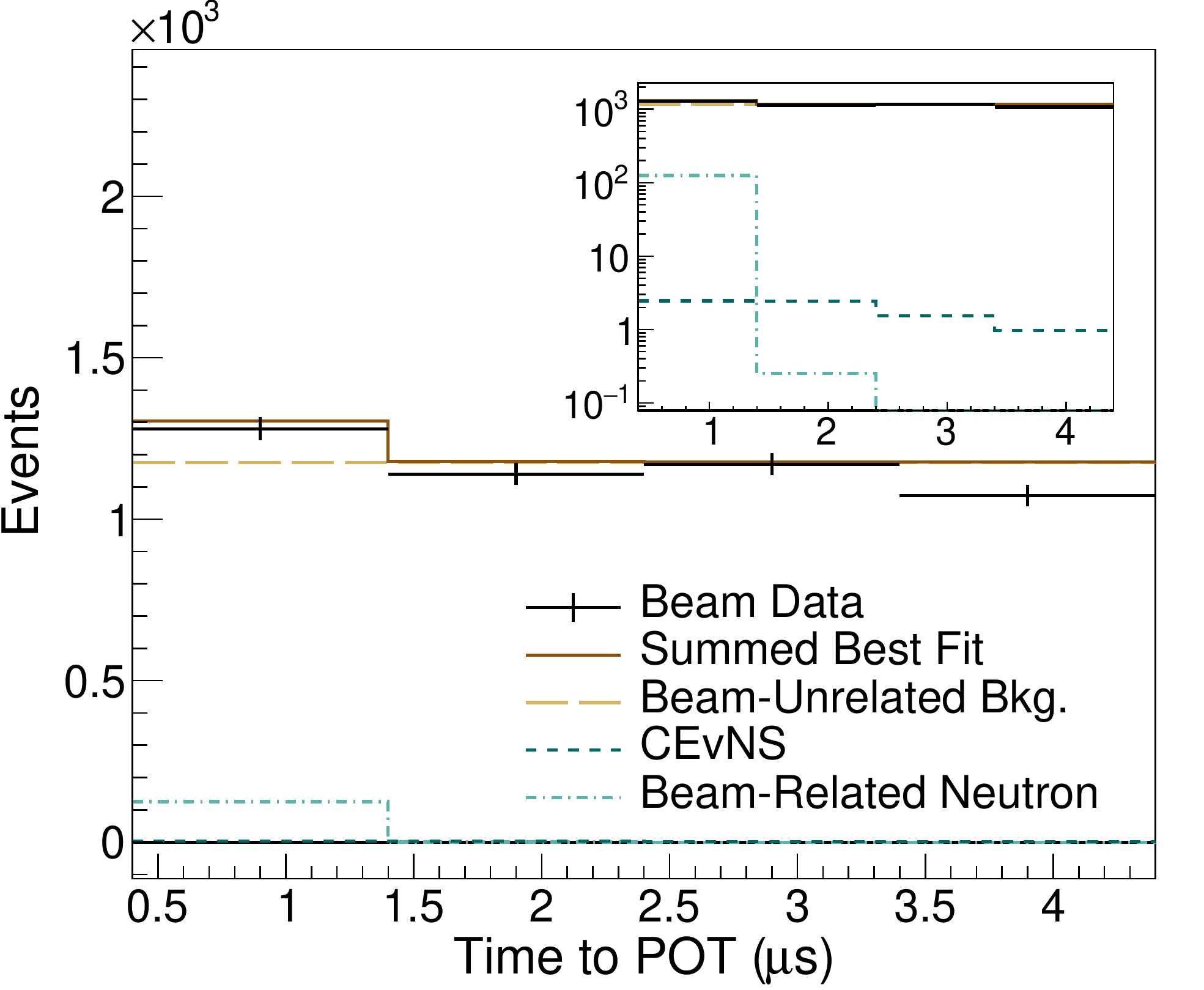}\\
        \centerline{(a)}
    \end{minipage}
    \begin{minipage}{0.325\textwidth}
        \includegraphics[width=\textwidth,page=2]{figures/likelihood_bf.pdf}\\
        \centerline{(b)}
    \end{minipage}
    \begin{minipage}{0.325\textwidth}
        \includegraphics[width=\textwidth,page=3]{figures/likelihood_bf.pdf}\\
        \centerline{(c)}
    \end{minipage}
      \caption{Projections of likelihood best-fit solutions together with the data in (a) time, (b) $F_{prompt}$, and (c) reconstructed energy. The \cevns curve shown is from the \SI{68}{\percent} confidence limit found. Inlaid plots show the spectra in log-scale to make the small contributions from the predicted \cevns distribution more visible.}
      \label{fig:likelihoodbfspectra}
\end{figure*}

The likelihood fit was performed by passing a total of 4663 events surviving the likelihood cuts to a 3D likelihood function in energy, time, and \fprompt space including beam-unrelated and BRN backgrounds along with a \cevns signal.  A profile likelihood curve was calculated as a function of the number of \cevns events and a frequentist confidence limit (C.L.) method~\cite{Feldman:1997qc, Thornton:2017etu, bib:mb}, along with a simple treatment of the large systematic errors, was used to place on upper limit on the number of \cevns events of \SI{<7.4}{events}. This result can be used to place a \SI{68}{\percent} C.L. on the stopped-pion flux-averaged cross section of \SI{<3.4e-39}{\cm\squared}, within twice the Standard Model prediction of \SI{1.8e-39}{\cm\squared}~\cite{Akimov:2018ghi}. These results are summarized in Table~\ref{tab:evtrates} and the projections in time, $F_{prompt}$, and reconstructed energy can be seen in Fig.~\ref{fig:likelihoodbfspectra}.

\begin{table}
\caption{Results of a maximum likelihood fit to the data (details in text). The quoted beam-unrelated background counts includes the statistical uncertainty in its determination from the strobe trigger sample.}
\label{tab:evtrates}
\begin{tabular}{p{1.8in} r}
\hline
sample size  & \num{4663}  \\
\hline
beam-unrelated background  & \num{4700 \pm 34} \\
fit BRN & \num[parse-numbers=false]{126 \pm 18(stat.) \pm 28(syst)} \\
1$\sigma$ (\SI{68}{\percent} C.L.) \cevns events & $<\num{7.4}$ \\
1$\sigma$ cross section & $<\SI{3.4e-39}{\cm\squared}$ \\
1$\sigma$ cross section sensitivity & $<\SI{7.1e-39}{\cm\squared}$ \\
\hline
\end{tabular}
\end{table}

\begin{figure}
    \centerline{\includegraphics[width=0.99\columnwidth]{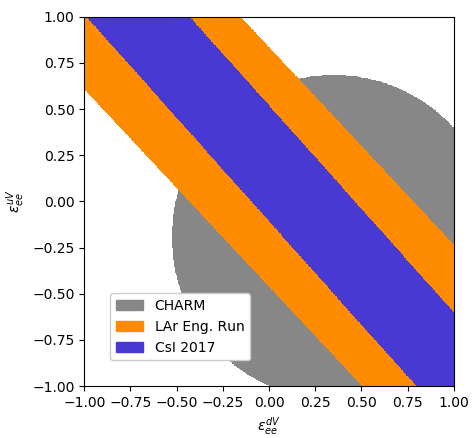}}
    \caption{\SI{90}{\percent} CL on NSI parameters $\epsilon_{ee}^{uV}$ and $\epsilon_{ee}^{dV}$ from this \cenns engineering run. The earlier \csi result~\cite{Akimov:2017ade} is confirmed and much of the pre-COHERENT phase space allowed by CHARM~\cite{bib:charm} is ruled out.}
    \label{fig:nsi}
\end{figure}

Using the same frequentist method a \SI{90}{\percent} C.L. on the cross section of $\SI{<8.3e-39}{\cm\squared}$ was extracted and used to set limits on the NSI couplings $\epsilon_{ee}^{uV}, \epsilon_{ee}^{dV}$~\cite{bib:barrancoCrossSection}. Under the assumption of heavy mediators, these couplings result in an overall scaling factor to the \cevns cross section~\cite{Akimov:2017ade}.  Fig.~\ref{fig:nsi} indicates the allowed parameter values consistent with this  \SI{90}{\percent} C.L. cross section.

\section{\label{sec:concl} Conclusions}
In this first result from the \cenns liquid argon detector as part of the COHERENT experiment, a dataset taken as part of an engineering run corresponding to \num{4.2e22} protons on the SNS target collected from Feb. 24, 2017 to May 28, 2017, has been analyzed.  The energy threshold in this configuration is not adequate for high sensitivity to \cevns[].  However, beam-related neutrons were characterized, further refining constraints on this important background which will inform future measurements. In addition, no BRN were observed in the delayed time window, outside of the beam pulse, consistent with previous measurements.  The observation of no significant beam excess does allow for a first limit on the \cevns cross section on argon within twice the SM prediction and for a corresponding limit on NSI. 

The \cenns detector was upgraded in the summer of 2017 to improve light collection and lower the energy threshold to \SI{20}{\keV nr}, and additional shielding was installed to minimize the dominant beam-unrelated background in Neutrino Alley. \cenns has collected \SI{>6}{\giga\watt hr} of data in this configuration with the sensitivity to make a first observation of \cevns on argon.
COHERENT is also working towards \cevns measurements with a \SI{2}{\tonne} \ce{NaI} array, also sensitive to charged current interactions, as well as with \SI{16}{\kg} p-type point-contact \ce{Ge} to maximize the neutrino physics capabilities at the SNS~\cite{Akimov:2018ghi}.

\section{Acknowledgments}
The COHERENT collaboration acknowledges the generous resources provided by the ORNL Spallation Neutron Source and thanks Fermilab for the continuing loan of the \cenns detector. We also acknowledge support from: the Alfred P. Sloan Foundation, the Consortium for Nonproliferation Enabling Capabilities, the Institute for Basic Science (Korea, grant No. IBS-R017-G1-2019-a00), the National Science Foundation,  the Russian Foundation for Basic Research (proj.\# 17-02-01077 A), and the U.S. Department of Energy Office of Science. Laboratory Directed Research and Development funds from ORNL and LLNL also supported this project.  This research used the Oak Ridge Leadership Computing Facility, which is a DOE Office of Science User Facility. 

\bibliography{lar_engrun_main.bib}
\end{document}